\documentclass[%
 reprint,
 amsmath,amssymb,
 aps, superscriptaddress, twocolumn
]{revtex4-2}
\usepackage{graphicx}
\usepackage{dcolumn}
\usepackage{bm}
\usepackage{amsmath}
\usepackage{amssymb}
\usepackage{CJK}
\usepackage{color}
\usepackage[pagewise,columnwise]{lineno}
\usepackage{hyperref}
\usepackage{epstopdf}
\usepackage{lipsum}

\begin{document}

\preprint{APS/123-QED}

\title{Revisiting the dynamics of Bose-Einstein condensates in a double well by deep learning with a hybrid network}

\author{Shurui Li}
\author{Jianqin Xu}
\author{Jing Qian}%
 \email{Corresponding author. E-mail: jqian1982@gmail.com}
\affiliation{Department of Physics, School of Physics and Electronic Science, East China Normal University, Shanghai 200062, People's Republic of China
}%
\author{Weiping Zhang}
\affiliation{School of Physics and Astronomy, and Tsung-Dao Lee Institute, Shanghai Jiao Tong University, Shanghai 200240, People's Republic of China}
\affiliation{Shanghai Research Center for Quantum Sciences, Shanghai 201315, People's Republic of China}
\affiliation{Collaborative Innovation Center of Extreme Optics, Shanxi University, Taiyuan, Shanxi 030006, People's Republic of China}

\date{\today}

\begin{abstract}
Deep learning, accounting for the use of an elaborate neural network, has recently been developed as an efficient and powerful tool to solve diverse problems in physics and other sciences. In the present work, we propose a novel learning method based on a hybrid network integrating two different kinds of neural networks: Long Short-Term Memory(LSTM) and Deep Residual Network(ResNet), in order to overcome the difficulty met in numerically simulating strongly-oscillating dynamical evolutions of physical systems. By taking the dynamics of Bose-Einstein condensates in a double-well potential as an example, we show that our new method makes a high efficient pre-learning and a high-fidelity prediction about the whole dynamics. This benefits from the advantage of the combination of the LSTM and the ResNet and is  impossibly achieved by a single network in the case of direct learning. Our method can be applied for simulating complex cooperative dynamics in a system with fast multiple-frequency oscillations with the aid of auxiliary spectrum analysis.


\end{abstract}

\maketitle


\section{Introduction}

Deep learning (DL) has been devoted to solving physical problems in recent years. With the help of its extraordinary abilities to make predictions about data and to extract meaningful features, exciting progresses have been made in the research fields of quantum dynamics  \cite{PhysRevLett.124.140502,PhysRevLett.124.113202}, many-body physics \cite{Carleo602}, precision measurement\cite{PhysRevLett.104.063603}, condensed-matter physics \cite{PhysRevLett.120.066401,Torlai2018,PhysRevX.7.031038,vanNieuwenburg2017,Carrasquilla2017} and quantum machine learning \cite{qaml, dlm, qeml}. When applying DL the primary problem faced by us is to establish a suitable network for the training data set. So far several neural networks(NNs) are utilized. For example, the earlier artificial neural network can approximate any distributions with only two neural layers \cite{ANN}; but it becomes inaccurate when the layer number increases \cite{backANN}. The restricted Boltzmann machine is an unsupervised method, which can uncover the information underlying the data by learning the joint distribution with Bayesian theory \cite{PhysRevE.96.022131}. It is difficult to train the system without a fine partition-function solution \cite{backRBM}. Other NNs such as the convolutional NN \cite{1998Gradient} and the recurrent NN \cite{RNN} are more universal in extracting local features of data and learning the relationships among elements with a high accuracy \cite{PhysRevD.101.023515, PhysRevResearch.2.023155}, despite that they may face the problem of gradient exploding or gradient vanishing due to the complexity of dynamics. Therefore a single type of NN becomes incapable of solving complex population dynamics.

The quality of simulating a fast-oscillating dynamics usually depends on the precision of sampling \cite{sample}. When the number of sampling points is insufficient, exact numerical ways may arise poor outcomes that are largely deviated from its real values \cite{Nyqui}. However a high-precision sampling will add to the costs of experiments; at the same time it also leads to a long-time calculation limited by the capacity of computers. Therefore it is quite difficult to accurately obtain the data of fast-oscillating dynamics both in theory and in experiment.
In the present work we propose a hybrid NN involving two different networks which are so-called
Long short-term memory(LSTM)\cite{1997Long} and Residual network(ResNet)\cite{7780459, SR}, in order to study the fast-oscillating dynamical evolution of systems.
Taking a generalized double-well dynamics of Bose-Einstein condensates(BECs) as an example, we show that our modified DL method
can well solve the time evolution of population difference between two wells with high fidelity of the phase and the frequency of oscillations. This modified DL refers to one kind of series connection between LSTM and ResNet with its performance enhanced by a pre-learning of periodicity in dynamics. ResNet will use the learning outcomes from LSTM to work better. In addition, an improved integrated DL based on modified DL is also proposed to facilitate the training time, which is enabled by changing the loss function for feedback. With this feedback mechanism, LSTM can benefit from ResNet, resulting in the reduction of data amount needed for training. 


For atomic Bose-Einstein condensates(BECs) trapped in a double-well potential, a large number of achievements were carried out arising a well-solved theoretical model \cite{PhysRevLett.118.230403, PhysRevA.98.063632, PhysRevA.98.043624, poland, Science.350.6267}. This leads to an easier generation of enough training data via the Runge-Kutta method beforehand. Our hybrid network enables a highly-efficient learning of frequency based on the periodicity verification of strongly-oscillating population dynamics. Moreover, even if the original data is not enough the LSTM network can also perform a reliable prediction on the higher frequency values of the population oscillating in the macroscopic quantum self-trapping (MQST) regime, which are far beyond the trained parameters. However, the numerical method fails to work in this case because of the inadequate sampling. 
Compared to the direct DL that only contains single kind of ResNet, integrating LSTM and ResNet can deeply increase the learning efficiency despite expending a slightly longer training time, since more layers are required for constructing this integrated network. The final infidelity obtained by estimating the deviation of fractional population difference can be much lower than 0.01 with the help of a hybrid network learning.

\section{Residual Network}

\begin{figure}
\centering
\includegraphics[width=8.7cm, height=3.0cm]{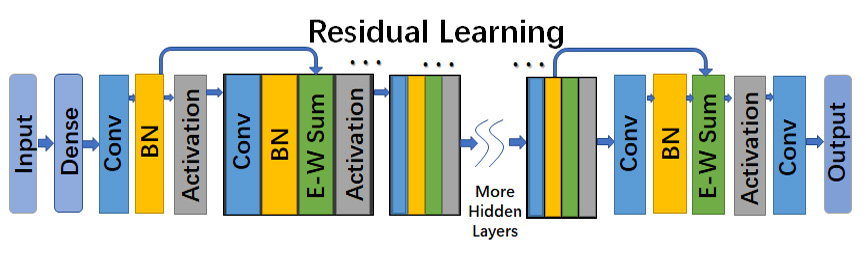}
\caption{Schematic diagram for ResNet involving a series of identically-distributed one-dimension convolutional (Conv) layers and batch normalization (BN) layers. At each convolutional layer it has a dilation rate except the first and the last layers. Different from the original convolutional NN scheme, arrows acting on Element-wise sum layers(E-W sum) can preserve the former hidden layer's information which facilitate the performance of DL.}  
\label{figure1}
\end{figure}

Structure of our residual network is shown by a series of blocks in  Fig. \ref{figure1}. The filter kernel serving as a central element in the learning process, should be carefully chosen.
We use filter kernels to scan matrice which enables the feature extraction from objects. By element-wisely multiplying matrice and filter kernels, machine can well learn the features of input data. However, with deeper networks, a traditional convolutional NN becomes more difficult to train, since we may face the problem of gradient vanishing or gradient exploding, which leads to a poor performance in DL. Hence, we choose a deep residual network(or so-called {\it ResNet})
to solve this problem. An element-wise sum in the ResNet labeled by arrows in Fig. \ref{figure1}, will prevent the loss of information from passing through deep networks, and also avoid the vanishing of gradient by adding the former input with present hidden layer's output. This simple process preserves the former information which means it increases the volume of data handled by the next layer. The detail of this technology can be seen in reference \cite{7780459}.
To be more specific, arrows linking the batch normalization layer and the element-wise sum layer in Fig.{\ref{figure1}}, point to the process of residual learning.  

From physical review, we are more interested in training or generating useful information within a ``black box" (the red frame in Fig.{\ref{figure5}b}) if the input is determined. Although training a high-performance network needs a lot of expense yet it may enable a robust prediction about the unknown results in a short-time period, especially while dealing with a complex system in the absence of analytical solutions. Our motivation is building an efficient learning protocol which can bridge the results from original numerical simulation and unknown prospective result based on the given input parameters. That may facilitate an easy and robust simulation for solving complex physical problems.

\section{Double-well dynamics with two BECs}

A double-well potential serving as a basic platform in the study of microscopic mediums, especially for atomic BECs \cite{bec1, bec2, bec3, bec4, bec5}, has been focused by researchers for decades. A typical double-well potential can be easily created by the combination of a periodic optical potential with strong harmonic confinement, played by a periodic one-dimensional light shift \cite{PhysRevLett.95.010402}.
If atomic BECs are confined in such a double-well optical trap, the effects of boson Josephson junction (BJJ) or MQST can be observed.

Our study starts from a brief introduction to the dynamical evolution of BECs in a double-well potential. The wave function $\psi(x,t)$ for two weakly-coupled atomic condensates is expanded as
\begin{equation}
\begin{aligned}
 \psi(x,t)=a_1(t) \phi_1(r)+a_2(t)\phi_2(r).
 \end{aligned}
\end{equation}
under the mean-field approximation. To be universal, two wells are asymmetric and has an energy difference $\gamma$ between them. A standard treatment for solving the population dynamics between two wells, will adopt the nonlinear two-mode dynamical equations (here $\hbar=1$)
\begin{equation}
\begin{aligned}
 i\frac{d}{dt}\left(\begin{array}{c} a_1 \\ a_2 \end{array} \right)=\hat{H}\left(\begin{array}{c} a_1 \\a_2 \end{array} \right)
 \end{aligned}
 \label{equation2}
\end{equation}
with the Hamiltonian $\hat{H}$ given by
\begin{equation}
\begin{aligned}
 \hat{H}=\left(\begin{array}{cc} \gamma+U_1N_1 & -K \\ -K & U_2N_2 \end{array} \right)
 \end{aligned}
\end{equation}
and the amplitudes $a_j=\sqrt{N_j}e^{i\theta_j}$ ($j\in(1,2)$) with $N_j$, $\theta_j$ the number and the phase of particles in the $j$th trap. Here $N_T=N_1+N_2$ is a constant and $U_j$ is the trapped atomic interaction which is related to the atomic scattering length. $K$ is the inter-trap coupling coefficient \cite{PhysRevLett.118.230403}. Upon defining the phase difference $\phi=\theta_2-\theta_1$ and the fractional population difference $z=\frac{N_1-N_2}{N_T}$, equation ({\ref{equation2}}) can be re-organized as
\begin{equation}
\begin{aligned}
    &\dot{z}(t)=-\sqrt{1-z^2(t)}\sin[\phi(t)],\\
    &\dot{\phi}(t)=\Delta E+\Lambda z(t)+\frac{z(t)}{\sqrt{1-z^2(t)}}\cos[\phi(t)],
\end{aligned}
\label{equation4}
\end{equation}
in which we rescale the time by $2Kt \rightarrow t$($2K$ is the frequency unit). Other dimensionless parameters are 
\begin{equation}
\begin{aligned}
    &\Delta{E}=\gamma/2K+(U_1-U_2)N_T/4K,\\
    &\Lambda=(U_1+U_2){N_T}/4K.
\end{aligned}
\label{equation5}
\end{equation}
accordingly, which indicate the imbalanced energy of two wells as well as the atomic tunneling rate respectively.
Solving $z(t)$ and $\phi(t)$ essentially gives rise to the whole population dynamics of atomic BECs in a double-well system.

It is clearly shown that the pair of variables $z(t)$ and $\phi(t)$ are canonically conjugate satisfying $\dot{z}$=$-\frac{\partial{H}}{\partial{\phi}}$, $\dot{\phi}$=$\frac{\partial{H}}{\partial{z}}$. Based on that, the Hamiltonian $\hat{H}$ can be rewritten as
\begin{equation}
\begin{aligned}
    \hat{H}=\frac{\Lambda}{2}{z}^2+\Delta Ez-\sqrt{1-z^2}\cos\phi,
\end{aligned}
\end{equation}
By simply assuming $\gamma=0$(symmetric) and ensuring $U_1=U_2=U$, it leads to $\Delta E=0$, $\Lambda=U{N_T}/2K$. The initial energy starts from
\begin{equation}
\begin{aligned}
    H_0=\frac{\Lambda}{2}{z(0)}^2-\sqrt{1-{z(0)}^2}\cos\phi(0),
\end{aligned}
\label{equation7}
\end{equation}
which is conserved during the dynamical evolution. According to the initial phase difference $\phi(0)=0$ or $\pi$ we can obtain the zero-mode or $\pi$-phase mode individually. Here we focus on the former case. $z(0)\in[-1,1]$ stands for the initial population imbalance and is tunable. In experiment via the adjustment of $\Lambda\sim U/K$ that characterizes the relative strength between interaction and tunneling rate, we can observe two different regimes: BJJ and MQST. The former meets $\left\langle z(t)\right\rangle=0$ and $\left\langle \phi(t)\right\rangle=0$ around $z=0$ when $\Lambda$ is small. As increasing $\Lambda$ that exceeds a critical value $\Lambda_c$, the atomic population oscillation becomes self-trapped giving to $\left\langle z(t)\right\rangle\neq0$ on average. This behavior is also accompanied by an unbound and increasing phase that is winding with time.

The critical value between Josephson oscillation and self-trapping effects has been analytically solved at $H_0=1$, arising \cite{PhysRevLett.79.4950}:
\begin{equation}
   \Lambda_c=\frac{2+2\sqrt{1-{z(0)}^2}\cos\phi(0)}{{z(0)^2}}
   \label{Lambdac}
\end{equation}
which means if the initial energy increases exceeding $H_0=1$ the fractional population difference will reveal a significant change, corresponding to the critical point between the two regimes.

\begin{figure}[ht]
\centering
\includegraphics[width=8.7cm, height=4.9cm]{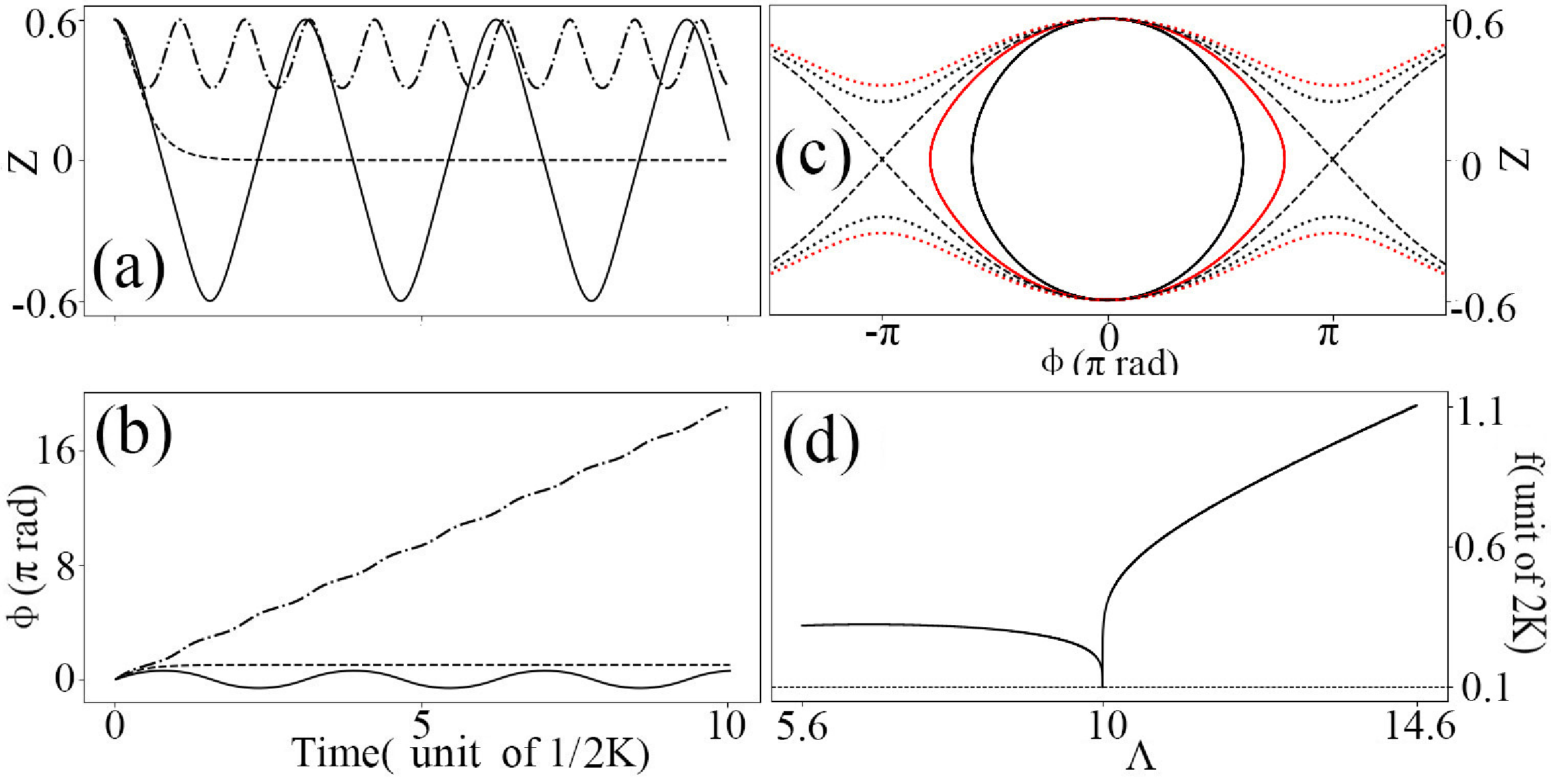}
\caption{(a-b) Numerical results $z(t)$ and $\phi(t)$ for $\Lambda = 6.2$(solid), 10(dashed, critical value) and 13.4(dash-dotted). (c) The ($z-\phi$) diagram under {$\Lambda$}= 6.2(black solid), 8.8(red solid), 10(black-dashed), 11.8(red-dotted), 13.4(black-dotted). (d) Single oscillating frequency $f$ of the population difference $z(t)$ for $\Lambda\in[5.6,14.6]$. At the critical point $\Lambda_c=10$, we set $f=0.1$.}
\label{theory}
\end{figure}

By using a standard fourth-order Runge-Kutta method, 
our numerical simulations for $z(t)$ and $\phi(t)$ are explicitly presented in Fig.\ref{theory}(a-b). As expected, both the population dynamics and the phase variation manifest the existence of 
a significant transition across the critical value $\Lambda_c$. If $\Lambda<\Lambda_c$ a BJJ appears between two wells accompanied by a small phase fluctuation in time, ensuring the average values $\langle z(t)\rangle=\langle\phi(t)\rangle=0$.
However when $\Lambda$ becomes larger, atom tends to be self-trapped in single-well with a time-dependent and increasing phase. Figure \ref{theory}(c-d) show a global representation of the phase diagram in the (z,$\phi$) space as well as the frequency $f$ of population dynamics versus $\Lambda$. The critical transition at $\Lambda_c = 10$ is clearly seen, agreeing with the discussions in (a-b). These numerical results allow us to establish 
a sufficient and accurate data set to train the neural network. In addition
the DL method enables a robust prediction about unknown results which are far beyond the test parameters in data set.

\section{Direct Learning}\label{dil}

To show the robustness of DL which can be applied for diverse physical problems, we first take this double-well model as an example, since both the fractional population difference $z(t)$ and the phase dynamics $\phi(t)$ of two BECs can be analytically solved here. 
Nowadays the DL method is regarded as a shortcut for this target. Because once a data set has been formed by using part of the original data obtained from numerical simulations, it reveals a strong ability of generating similar data $\phi(t)$ and $z(t)$ with the training network. Most interestingly, benefiting from the ability of strong prediction it can even give other unknown results which are beyond the original data set.

Our work starts from equation (\ref{equation4}) by setting {\it e.g.} $z(0)=0.6$, $\phi(0)=0$, arising $\Lambda_c=10$. The original parameter range is fixed to $\Lambda\in[5.6,14.6]$ covering two regimes: MQST and BJJ. The data set is formed by uniformly separating 90001 $\Lambda$ values within this range and solving $\phi(t)$ and $z(t)$ for each $\Lambda$ via the method of fourth-order Runge-Kutta. The time interval is $\Delta t=0.02$ with a unit of $1/2K$. As a result, we can finally obtain two two-dimension data sets for $z$ and $\phi$ in the space of $(\Lambda,t)$. Each contains 90001$\times$501 datas.
While training the network with data sets we randomly choose 9000 samples as a validation set, leaving the remaining 81001 samples for training. All computing results are based on Intel Core i7-7900X @3.30GHz and GeForce RTX 2060 with 1024 batch size and 32 filter size on Python 3.8.

\begin{table}
\centering
\begin{tabular}{l|l|l|l|l}
\hline
Layers & Loss for Tanh & Loss for Sine   & Time(s) & Efficiency \\
\hline

5      & 0.0173        & 0.0192 & 7       & 8.26   \\

\textbf{7}      & \textbf{0.0112}        & \textbf{0.0131} & \textbf{9}       & \textbf{9.92} \\

9      & 0.0097        & 0.0102 & 11      & 9.37     \\

11     & 0.0088        & 0.0092 & 15      & 7.58   \\

13     & 0.0086        & 0.0089 & 20      & 5.81    \\
\hline
\end{tabular}
\caption{\label{tab:table1} Comparison for the training efficiencies under different number of hidden layers, activation functions and the average spending time for one epoch. Efficiency is given by the speed(=1/Time) divided by the Tanh's loss.}
\end{table}

An efficient learning depends on the design of network. To search for a suitable network, in Table {\ref{tab:table1}} we select different hidden-layers and calculate the losses after 400 epochs by using different activation functions. Here the ``layers" means the number of hidden convolutional layers.
In general as the increase of layer numbers the Mean-squared-error(MSE) becomes lower, however at the expense of a longer training time. To propose a low-expense method with high precision, we use {\it efficiency}(the last column in Table {\ref{tab:table1}}) as an indication for one network's compromised performance, which is 
defined by the training speed divided by the Tanh-type loss. It is clearly shown that a seven-layer model is the best one because of the maximal efficiency value $\sim 9.92$(bold). Also, we compare the results based on two different activation functions: Tanh and Sine. The Tanh-type function can bring a lower learning loss than the Sine function, which will be considered in the calculation.
In a practical training, we choose optimizer Adam for implementing the optimization \cite{2014Adam}. In order to prevent over-fitting, the decay factor is 0.8, the minimum learning rate is $5\times 10^{-6}$ and the patience is 2.0 in the stage of 
Early-stopping and adaptive learning. 
The MSE loss function $\mathcal{L}$ is defined by
\begin{equation}
    \mathcal{L}=\frac{1}{N_{\Lambda}}\sum_{\Lambda}^{N_{\Lambda}}\sum_i^{N} \frac{(z(\Lambda,t_i)-g(\Lambda,t_i))^2}{N}
\label{equation9}
\end{equation}
 where {$z(\Lambda,t_i)$} is the learning outcome {\it i.e.} the fractional population imbalance at $t_i=\Delta t\times i$($i\in[1,N]$) for each $\Lambda$. $g(\Lambda,t_i)$ is the ground-truth from numerical solutions for each $\Lambda$. $N$ is the length of arrays and $N_{\Lambda}$ is the number of $\Lambda$ learned in ResNet. Here $N=501$, $\Delta t=0.02$, $N_{\Lambda} = 81001$.

\begin{figure}
\centering
\includegraphics[width=8.7cm, height=4.3cm]{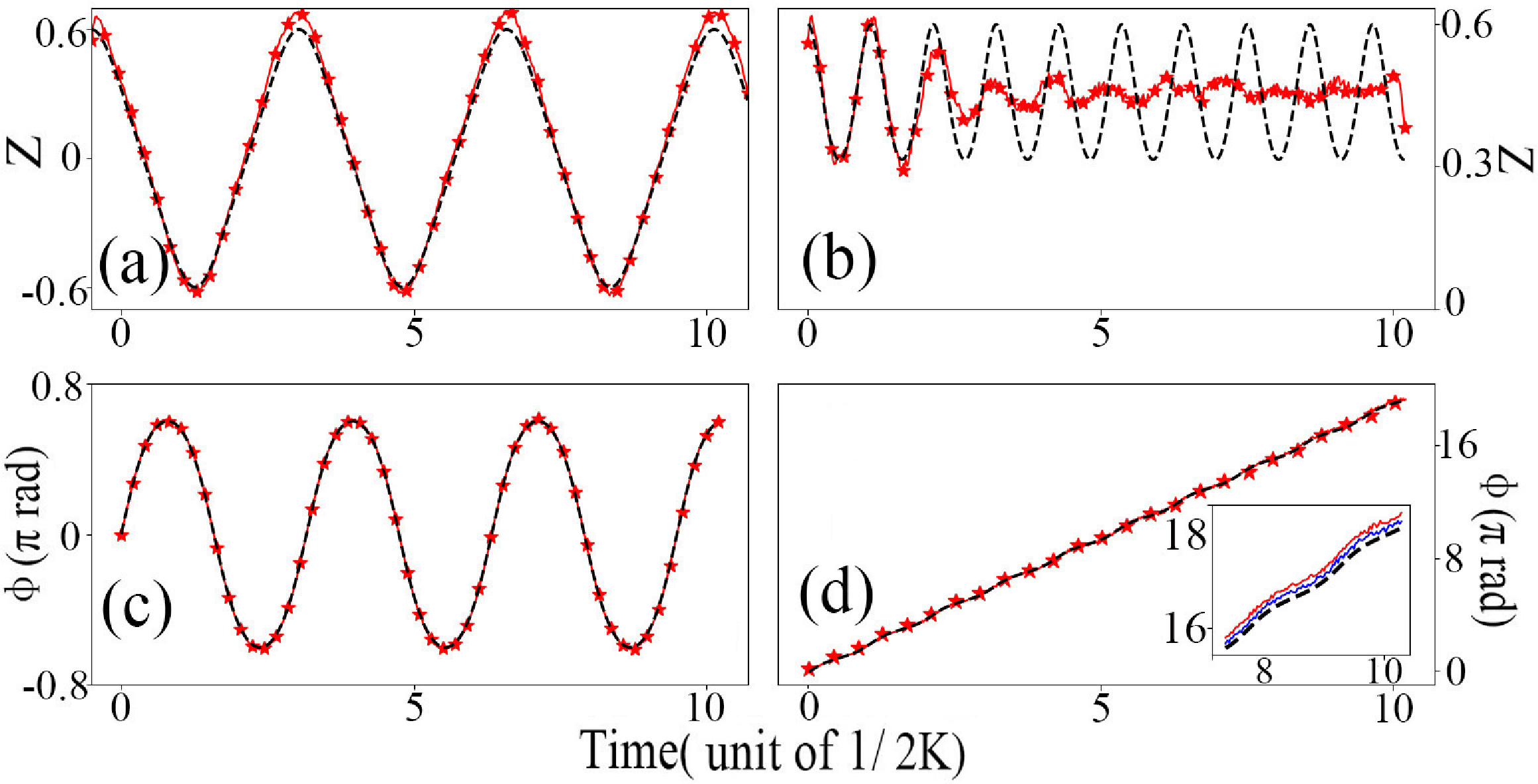}
\caption{Direct DL of $z(t)$[(a-b)] and $\phi(t)$[(c-d)] marked by red stars. Numerical results are contrastively given by the black-dashed curves, as similar as the results shown in Fig.2(a-b).
For (a,c) $\Lambda=6.2$(BJJ regime), $z(t)$ and $\phi(t)$ solved from the two methods have a perfect agreement. As turning to the regime of MQST, {\it i.e.} (b,d) $\Lambda=13.4$, only $\phi(t)$ well agrees with the numerical prediction and $z(t)$ exhibits a larger mismatch as time increases. Inset of (d) shows the behavior of $\phi(t)$ under different layer numbers.}
\label{figure2}
\end{figure}

Figure \ref{figure2} globally displays the population difference $z(t)$ and the phase variation $\phi(t)$ in two regimes, extracted from direct DL and numerical ways respectively. As for the learning of phase variation our results well fit with the numerical predictions see (c-d).
Note that when plotting (d) we do not constrain the range of $\phi$. So due to the linear increasing of $\phi$ accompanied by a slight oscillation in the long-time regime, the number of layers(here we choose seven hidden layers) becomes insufficient. As shown in the inset of (d), with the increase of layer numbers: {\it e.g.} seven layers(red-solid), nine layers(blue-solid), a nine-layer network in direct DL provides a better agreement with the numerical result(black-dashed).
However, the direct DL with single kind of network fails to work well in the case of $z's$ learning, especially in the MQST regime. Because in this regime two atomic BECs trapped in one well, will show a fast-oscillating behavior, leading to the breakdown of our thin single network(ResNet). We find the learning outcomes can not catch up with the change of $z(t)$, see (b). In the core of direct DL, such a rapid oscillation with small peak-peak amplitudes can not be easily trained by a limited database. Because the small loss in the whole process would cause gradient vanishing, making the whole learning process inefficient.

To improve the learning quality, especially in the case of MQST where the dynamics of $z(t)$ behaves as rapid oscillations with $\langle z(t)\rangle\neq0$, we have to search for a new learning method. Since the periodicity of population dynamics is apparently seen, it is easy to decompose the function into two components: frequency and amplitude within one period. So we can learn them separately based on the dynamical character in one period, instead of implementing a global optimization for all hyper-parameters. This new learning method will be beneficial owing to the use of a low-expense network with fewer number of hidden layers as compared to the direct DL, if same accuracy is attainable. The modified network only needs a small receptive filed size which indeed reduces the number of kernel layers and its size. Before carrying out our new method we have to verify the periodic feature of $z(t)$.

\section{Modified Learning}\label{mol}
\subsection{Periodicity verification}
Periodicity verification can facilitate the design of network in modified DL. Compared to the traditional ways for that purpose such as
Fourier Transform(FT) \cite{acfft} and Auto-correlation Function(ACF) \cite{acfft}, here we choose the method originating from the chaotic physics to determine $z(t)$'s periodicity since it is more visible. This is called \textit{Poincar\'{e} section}  \cite{P2, P3}. For an orbit in a two-dimension phase space, one could always select an appropriate hyperplane which should not contain or be tangent with this orbit. In general, if there exists a fixed-point or fixed number of discrete points, the orbit must have a determined period \cite{liuzh}. Based on the nonlinearity of equation (\ref{equation4}) we use this method to determine the periodicity of dynamics.

In this double-well system, we choose {$\phi$}=0 as a hyperplane in the BJJ regime and {$\phi$}={$\pi$} in the MQST regime. We draw the points where the trajectory of ({$z-\dot{z}$}) crosses this hyperplane. Specific examples are displayed in Fig.\ref{figureA1} of Appendix A, in which we compare the relations of $(z-\phi)$, $(z-\dot{z})$ and the Poincar\'{e} maps under different $\Lambda$ values. When plotting the Poincar\'{e} maps
the periodic character of $z(t)$ can be clearly confirmed by the presence of some discrete points.


\subsection{Frequency learning}

A direct simulation for the frequency $f$ as a function of $\Lambda$ has been solved by adopting numerical ways in Fig.\ref{theory}d, where the critical frequency at $\Lambda=\Lambda_c$ is fixed to be 0.1. When turning to the problem of solving an one-dimensional array of data, {\it e.g.} learning frequency of fractional population difference $z(t)$, ResNet is not the best choice since it typically suits for extracting features of a two-dimensional array \cite{d21,d22}, such as learning $z(\Lambda,t)$ and $\phi(\Lambda,t)$ in direct DL. Here we introduce a modified Recurrent Neural network which is so-called LSTM for that purpose \cite{1997Long}. This method has been applied for studying knot types of polymer conformations \cite{PhysRevE.101.022502} and phase-modulation stabilization in quantum key distribution \cite{PhysRevApplied.12.014059}.
LSTM is a model which could memorize the important part of the previous context in a data series and forget other irrelevant parts selectively. Therefore, it performs pretty well in time-series fitting and prediction due to the consideration of correlation in the context. Although its weakness also lies in the requirement for more time to train, in contrast to the case of using ResNet when same number of layers are required. Luckily it performs better in preventing from gradient vanishing and exploding in the long-range-series training.

\begin{table}
\centering
\begin{tabular}{l|l|l|l|l|l}
\hline
Layers & Tanh   & Sine   & Mixture & Time(s) & Efficiency  \\
\hline
4      & 0.0220  & 0.0217 & 0.0206  & 20      & 2.43   \\

\textbf{6}      & \textbf{0.0119} & \textbf{0.0103} & \textbf{0.0096}  & \textbf{25}      & \textbf{4.17}  \\

8      & 0.0103 & 0.0104 & 0.0081  & 32      & 3.86      \\

10     & 0.0097 & 0.0099 & 0.0079  & 41      & 3.09   \\

12     & 0.0094 & 0.0099 & 0.0078  & 50      & 2.56    \\
\hline
\end{tabular}
\caption{\label{table2}%
 Relevant parameters under the modified LSTM network. Way of {\it Mixture} means that half of layers use Tanh-function and the other half layers use Sine-function. The final efficiency is given by the speed divided by the Mixture's loss. Here layers point to the number of hidden LSTM layers.}
\end{table}

\begin{figure}
\centering
\includegraphics[width=8.7cm, height=3.8cm]{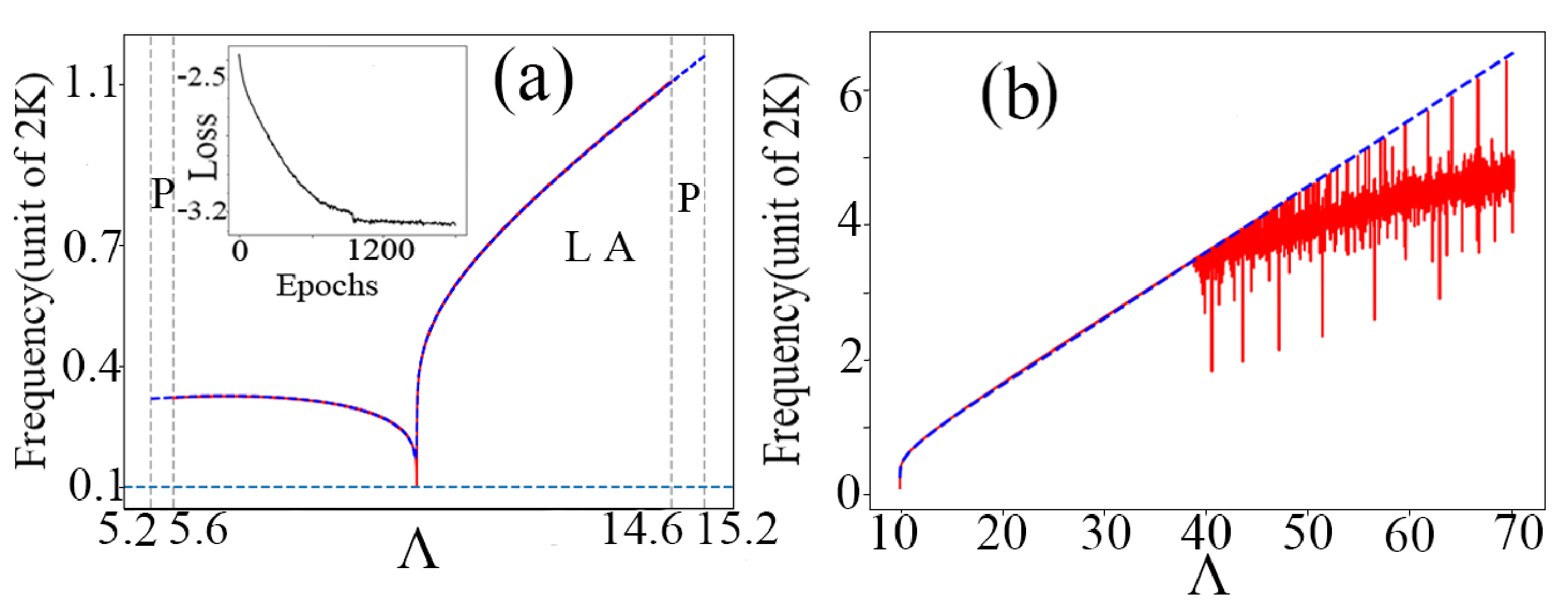}
\caption{(a) Frequency learning in the original learning area labeled by ``LA": $\Lambda\in[5.6,14.6]$. The prediction results denoted by ``P" are given in the area of $\Lambda\in[5.2,5.6]$ and $[14.6,15.2]$. Our LSTM learning outcomes(blue-dashed) well agree with the numerical results (red-solid). Insert shows the corresponding MSE loss in the form of logarithmic function versus the training epochs. (b) Frequency learning under an extensive range of $\Lambda\in[10,70]$. Here same linetypes are used.}
\label{figure4}
\end{figure}

The results of frequency learning are demonstrated in Table \ref{table2} and Figure \ref{figure4}. Table \ref{table2} compares the MSE losses under different hidden layers by using various activation functions: Tanh, Sine and Mixture.
For obtaining an optimal network we also consider the time expense, leading to the result of efficiency which can be used to quantitatively characterize the learning performance. We find only six hidden LSTM layers(bold) is sufficient for learning. Notice that here the definition of efficiency depends on the application of LSTM network which is different from that used in Table I for the case of direct DL. Besides we also use Early-stopping to avoid over-fitting. Root mean-square prop  is chosen as a new optimizer in LSTM network. The original training area is the same as in direct DL {\it i.e.} $\Lambda\in[5.6, 14.6]$ and the results from an extensive range of $\Lambda\in[5.2, 15.2]$ are also given based on the prediction of LSTM. A good agreement between our learning outcomes and the numerical results is apparently observed in Fig.\ref{figure4}a except for the critical point $\Lambda_c=10$ since the network treats it as an outlier(singularity) due to the discontinuity. Inset plots the change of MSE loss with epochs which has been reduced to be much smaller than 0.001 when epoch is larger than 1200. Therefore the way of LSTM is suitable for an individual frequency learning in the training process.

Remarkably, with the increase of $\Lambda$ the oscillation of dynamics becomes stronger. The traditional numerical method is unable to give accurate results due to the calculation precision limited by a sampling interval $\Delta t=0.02$. Reducing $\Delta t$ could enhance the precision yet adding to the cost. So a bigger mismatch will appear between the numerically-estimated values and the real numbers when $\Lambda\gtrsim 40$, see the red-solid curve in Fig.\ref{figure4}b. This numerical curve is obtained by computing the time duration between two same waves for extracting the information of frequency. It is apparent that the oscillating frequency becomes much larger as {$\Lambda$} increases which makes the fixed sampling interval inadequate. However the LSTM network can preserve a robust prediction for higher frequency values (blue-dashed, $\Lambda\in[10,70]$) under same initial preparation. Remember the initial data set based on $\Lambda\in[5.6,14.6]$ and $\Delta t=0.02$ are unchanged in the learning. Therefore the LSTM network is expected to be more suitable for solving the fast-oscillating behavior even if the sampling points are inadequate or in low-frequency regime.

\subsection{Modified network}\label{sec:C}

\begin{widetext}

\begin{figure}
\centering
\includegraphics[width=10cm, height=5.5cm]{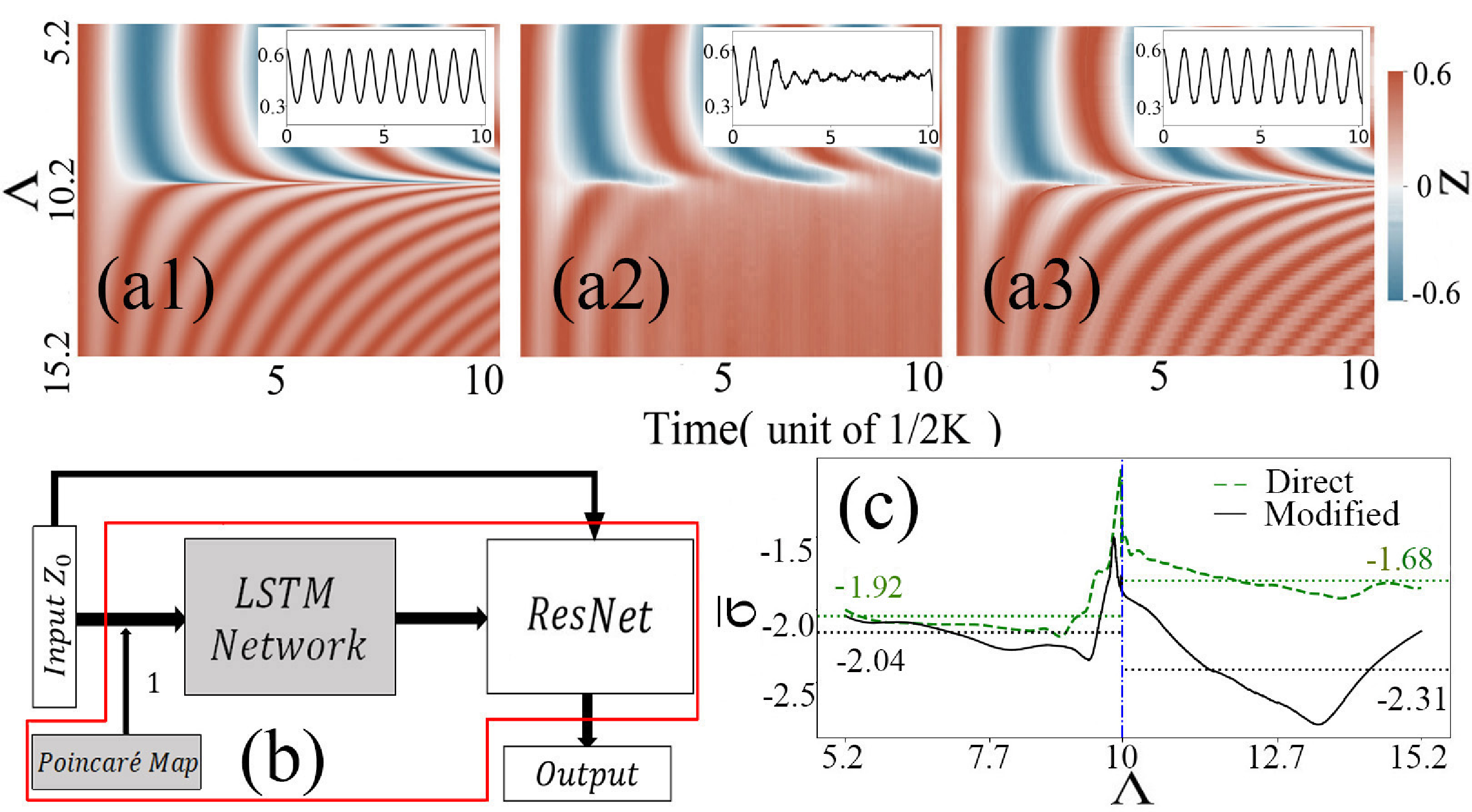}
\caption{Representation of the population dynamics $z$ in the space of $(t,\Lambda)$ under (a1) numerical simulation, (a2) direct DL, (a3) modified DL. Insets show the real population dynamics $z(t)$ for $\Lambda=13.4$ using different methods. (b) The procedure of our modified network by sequentially combining LSTM and ResNet. Labeling ``1" denotes the truth of dynamical periodicity verified by using \textit{Poincar\'{e} section}, as same as the labeling ``1" in Fig.6b. (c) Comparison of relative errors in logarithm form between direct DL(green dashed) and modified DL(black solid). Average values (dashed lines) are also given separately.}
\label{figure5}
\end{figure}

\end{widetext}

A new modified network can be reconstructed by taking account into the role of LSTM in the learning frequency. Thinner networks can only capture limited local information with a small receptive field size \cite{vnet}. Therefore, our method makes it possible to produce a hybrid network by combining LSTM and ResNet which is able to generate the whole population dynamics within a small receptive field size. In this hybrid learning network as shown in Fig.\ref{figure5}b, the former LSTM network is used for frequency learning by the way of Poincar\'{e} map at the first stage. Additionally, the ResNet is treated as a core network for generating the whole dynamics of the fractional population difference $z(t,\Lambda)$. For the ResNet learning, we also use the MSE loss function $\mathcal{L}$ [equation (\ref{equation9})], in which the parameters {$z(t_i)$}, {$N$}, {$g(t_i)$} are all obtained after implementing LSTM. In the core of our procedure [see red box in Fig.\ref{figure5}b], we use the Poincar\'{e} maps and LSTM network integrated to transmit extra information to ResNet. More importantly the way of Poincar\'{e} map can improve the learning fidelity via periodicity verification before learning. 

Figure \ref{figure5}(a1-a3) globally present the results of $z(t,\Lambda)$ under numerical simulation, direct DL as well as our modified network learning. A special case of $z(t,13.4)$ is shown in the insets accordingly. Clearly the direct DL method gives rise to poor results in the MQST regime as compared to the numerical calculation because of the strong oscillation. The direct DL only contains single ResNet that is insufficient for grasping the fast-oscillating behavior. However a significant improvement can be attainable if two blocks of Poincar\'{e} map and LSTM network are added in the procedure. 
A quantitative comparison between direct and modified DL results uses the mean relative error
\begin{equation}
   \bar{\sigma}(\Lambda)=\frac{1}{N}\sum_{i}^{N}\frac{|z(t_i, \Lambda)-g(t_i,\Lambda)|}{0.1+|g(t_i,\Lambda)|}
\label{rela}
\end{equation}
where $z(t_i,\Lambda)$ and $g(t_i,\Lambda)$ obtained from the network training and the numerical ground-truth, are also $\Lambda$-dependent. Here 0.1 serves as an arbitrary small quantity to overcome the divergency. The relative error {$\bar{\sigma}$} is a metric of the final performance standing for the derivation between learning outcome and numerical ground-truth, {\it i.e.} the infidelity of learning.
While the MSE loss is an indicator at the training stage.
In the BJJ region two learning ways lead to comparable errors ($10^{-1.92}$ and $10^{-2.04}$), confirming the efficiency of both learning methods. However as turning to the MQST region the modified DL apparently leads to the reduction of mean relative error from $10^{-1.68}$ to $10^{-2.31}$, which extremely increases the capability of network in learning a fast-oscillating dynamics.

 By now, the hybrid system that contains a sequential LSTM and ResNet networks can provide an improved learning for the population dynamics, especially in the MQST regime. While we also note that, two networks must be trained independently suffering from a long-time expense. A typical time expense is 34 seconds for one-epoch training. To overcome this weakness we propose an integrated network by combining LSTM and ResNet together with the help of feedback mechanism. This new learning method can further accelerate the training speed, saving the time by 15$\%$ or more in an one-epoch training.

\section{Integrated Learning}
\label{inl}

A universal study for the population dynamics in an asymmetric double-well is performed. Motivated by Ref.\cite{wy1} it is possible to utilize the feedback of loss from ResNet learning to train the former LSTM network, instead of a direct frequency learning by it, in order to  accelerate the learning speed. The role of LSTM is providing the frequency information for the next ResNet learning due to the thin and small convolutional network of Resnet. Similarly, the periodicity verification is carried out by using \textit{Poincar\'{e} section} as shown in Fig.\ref{figureA2}. The feedback mechanism in our integrated DL scheme means that the loss function of ResNet can be treated as an indicator for the LSTM network's learning, and it would largely improve its learning efficiency. Because the MSE loss function defines the deviation between the final outcome and the ground-truth value.

\begin{widetext}

\begin{figure}
\centering
\includegraphics[width=10cm, height=5.5cm]{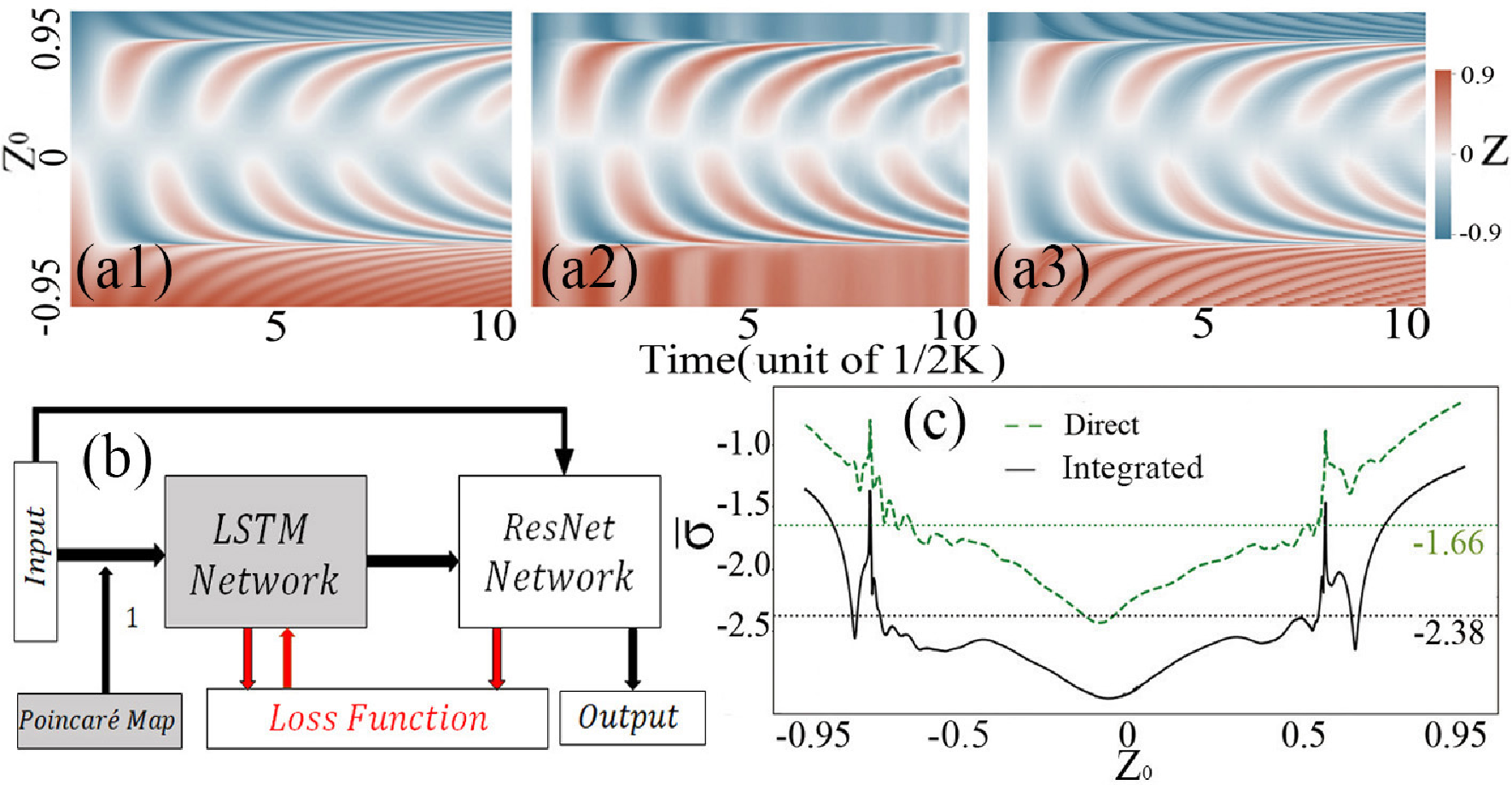}
\caption{(a1-a3) Representation of the population dynamics $z(t)$ in space of $(t,z_0)$ using the methods of numerical simulation, direct DL and integrated DL respectively. (b) Integrated DL procedure. The frequency learning by LSTM is facilitated due to the use of feedback mechanism from ResNet (red arrows). (c) Relative errors $\bar{\sigma}$ in logarithm form versus the initial imbalanced population $z_0=z(t=0)$ from direct DL (dashed line) and integrated DL (solid line).}
\label{figurenew}
\end{figure}

\end{widetext}

In the practical training we follow the procedure diagram in Fig.\ref{figurenew}b. It differs from the modified DL mainly by using a new loss function $\mathcal{L_I}$, which can be expressed as
\begin{equation}
\begin{aligned}
    &\mathcal{L_I}=\mathcal{L}-k\sum_{z_0}^{N_{z_0}^{\ast}}\frac{(f(z_0)-f_g(z_0))^2}{N_{z_0}^{\ast}},\\
\end{aligned}
\label{loss_eq}
\end{equation}
where $k$ is a weight factor and $N_{z_0}^{\ast}$ denotes the partial data number to train LSTM. In an asymmetric well we use the initial population difference $z_0$ as a tunable parameter, instead of $\Lambda$. The first term $\mathcal{L}$ standing for the loss estimation in single ResNet, takes a similar form like equation (\ref{equation9}). The second term serves as an auxiliary for showing the deviation of real frequency $f$ from the ground-truth $f_g$ with respect to each $z_0$. Therefore the ResNet learning can offer an efficient feedback for the frequency learning in the LSTM network.
Our scheme follows the procedure in Fig.\ref{figurenew}b. In order to minimize $\mathcal{L_I}$ as well as to speed up the convergence of this integrated network, we carry out an individual training for two networks at initial steps. After that we fix the ResNet and let the LSTM being trained for steps to minimize $\mathcal{L_I}$ accompanied by generating an array of frequency for each $z_0$. In addition we also simultaneously train ResNet to reduce the value of $\mathcal{L}$. Based on our practical training, we set four steps for LSTM's training and five steps for ResNet in each iteration, the weight factor is $k=1.2$.

The integrated DL bases on the fractional population difference $z(t,z_0)$ in an asymmetric double well where $\gamma=1.0$ is assumed. Here we set $\Lambda=8$ and treat $z_0=z(t=0)\in[-0.95,0.95]$ as a variable. After implementing a periodic verification of $z$ with the method of Poincar\'{e} maps (see Fig. \ref{figureA2} in Appendix A) we do the learning following the procedure of Fig.{\ref{figurenew}}b. Figure {\ref{figurenew}}(a1-a3) plot similar results as Fig.\ref{figure5}(a1-a3) yet within the $(t,z_0)$ space. 
Clearly as compared to the direct DL, the results obtained from the integrated DL in (a3) have a better agreement with the numerical results shown in (a1). To make this discussion more quantitative, we also calculate the relative errors $\bar{\sigma}$ of two learning methods. From Fig.{\ref{figurenew}}c, it is clearly shown that the integrated DL method performs better than the direct DL for arbitrary $z_0$ owing to its smaller relative errors. Besides we observe that $\bar{\sigma}$ will increase if $|z_0|$ is enhanced because of the increasing frequency.
 In the integrated DL we only have to compute partial data for estimating the frequency {\it i.e.} $N_{z_0}^{\ast}=1201$(the total number is $N_{z_0}=7601$), benefiting from the feedback mechanism. While it can reduce the relative error $\bar{\sigma}$ {\it i.e.} the infidelity, from $10^{-1.66}$ to $10^{-2.38}$ on average by using an integrated neural network.

\section{Conclusion and Outlook}

Before ending we compare the efficiencies of three DL methods. Note that here the efficiency is defined as the training speed divided by the relative error $\bar{\sigma}$, not the loss function. Given the original data are obtained by solving eq. (\ref{equation4}) we display the learning results of various methods in Table {\ref{tab:table3}}. The original data set for $z(t)$ is generated by setting $\Lambda\in[5.6,14.6]$. Although the training process costs much longer time than the generation it is possible to predict unknown results beyond the given parameter range. Note that here a direct DL with a total 13-layer ResNet fails to fit the real population dynamics on the MQST regime, leading to a poor efficiency $\sim 2.39$(the infidelity attains 0.0209 when $\Lambda\in[10,15.2]$).
However, accompanied by an auxiliary LSTM network that provides a certification of the oscillating frequencies, the resulting hybrid network(modified DL or integrated DL) has shown a significant improvement for the learning efficiency. The efficiency reaches as high as $\sim5.47$(the final infidelity obtained for $z(t)$ is far smaller than 0.01) for the integrated DL. Besides the hybrid network can also provide a precise learning in both BJJ and MQST regimes, and can even make convincing predictions about data with respect to the unknown parameters. When turning to the case of learning higher-frequency dynamics such as in the MQST regime, we find that the Runge-Kutta method becomes inaccessible due to the finite sampling interval. Instead, our LSTM network reveals an ability of making predictions about higher frequency values based on the same data set.

\begin{widetext}
\begin{table*}
\centering
\begin{tabular}{l|l|l|l|l|l}
\hline
Method & Training(s) & Generation(ms)   & Parameter $\Lambda$   & Relative error $\bar{\sigma}$ & Efficiency \\
\hline
Runge-Kutta   & -   & 1.15  & [5.6, 14,6] & - & -  \\

Direct Learning  & 20    & 3.98 & [5.2, 10] & 0.0123 & 4.07\\

& 20 & 3.98 & [10, 15.2] & 0.0209 & 2.39     \\

Modified Learning  & 34    & 5.37 & [5.2, 15.2] & 0.0066 & 4.46  \\

Integrated Learning  & 29    & 5.19 & [5.2, 15.2] & 0.0063 & \textbf{5.47}  \\
\hline
\end{tabular}
\caption{\label{tab:table3} Comparison for time costs and efficiencies of four methods in the training and generation of data $z(t,\Lambda)$ in one epoch. Taking the dynamics of BECs in a symmetric double-well potential as an example, the training parameter range, the relative error and the learning efficiency are comparably given under a same number of training layers. The total layers are same for modified DL and integrated DL which include 6 hidden LSTM layers and 7 convolutional ResNet layers. For direct DL we use a total 13-layer ResNet.}
\end{table*}
\end{widetext}

In conclusion we develop a hybrid ``{\it LSTM+ResNet}" network to revisit the population and phase dynamics of two BECs in a double-well potential. Because the fast-oscillating behavior of population difference in the MQST regime makes the traditional direct DL inaccessible,
we propose a novel learning method which integrates LSTM and ResNet networks and then brings about a strong promotion to the learning efficiency. The LSTM also represents a robust prediction for the characteristic periodicities of data sets. This hybrid network performs better with a high precision and a very low relative error (the final infidelity obtained for the fractional population difference) $\sim0.006$.
Most importantly, it shows a strong ability of prediction for the data in the fast-oscillating regime which are far beyond the original data set given by the exact numerical calculations in the slow-oscillating regime. This prediction capability of method could find more applications in solving problems of physics and other sciences, which are lack of sufficient parameters restrained by the experimental conditions or costs.

As an outlook we discuss how our hybrid network could work for other systems possessing more complex population dynamics which can not be analytically solved. For example, when the effective three-body interactions of BECs are not negligible there is no longer an analytical solution of the double-well dynamics and the evolution of population difference possesses complex multi-frequency oscillations \cite{PhysRevA.97.013609}. Then a Fourier frequency analysis combined with two algorithms, {\it Attention} and {\it Sparsity} can help us to realize efficient learning. If there exists several frequency components from the Fourier transformed data, {\it Attention}, which is a method in computer science to assign weights for values, will learn the weight of each frequency in the multi-frequency oscillating dynamics. However, if the exact frequencies are unknown, we have to put a series of candidates into the set.  Thus, the frequency set may have many irrelevant elements, making 
the learning outcome polluted. Serving as an enhancement trick for the Attention mechanism, {\it Sparsity} ensures the sparsity of used frequencies, which can make the selected frequencies more effective
and reduce the complexity and time expense for learning \cite{sparsity}. In short, with the aid of {\it Attention}, we can obtain the weight of each frequency component and reconstruct them into the final outcome where {\it Sparsity} may help learning in a continuous parameter system. Finally, 
remember that our DL method depends on a data set to train. This data set can be generated by solving theoretical equations. Even if the problem is analytically unsolvable the sample data can still be collected from experimental measurements. So our method not only applies to theoretical simulations but also to experimental estimation. Moreover, our integrated DL method is not restricted to a sufficient sampling which means an extensive application in various systems is desirable in the future.

\begin{acknowledgments}
This work was supported by the NSFC under Grants No.11474094 and No.11104076, No.11654005; the Science and Technology Commission of Shanghai Municipality under Grant No.18ZR1412800; the National Key Research and Development Program of China under Grant No. 2016YFA0302001; the Shanghai Municipal Science and Technology Major Project under Grant No. 2019SHZDZX01, the Shanghai talent program.
\end{acknowledgments}

\newpage
\appendix

\section{Numerical results of \textit{Poincar\'{e} section}}
\label{appendix}

\begin{figure}[h]
\centering
\includegraphics[width=8.2cm, height=8.2cm]{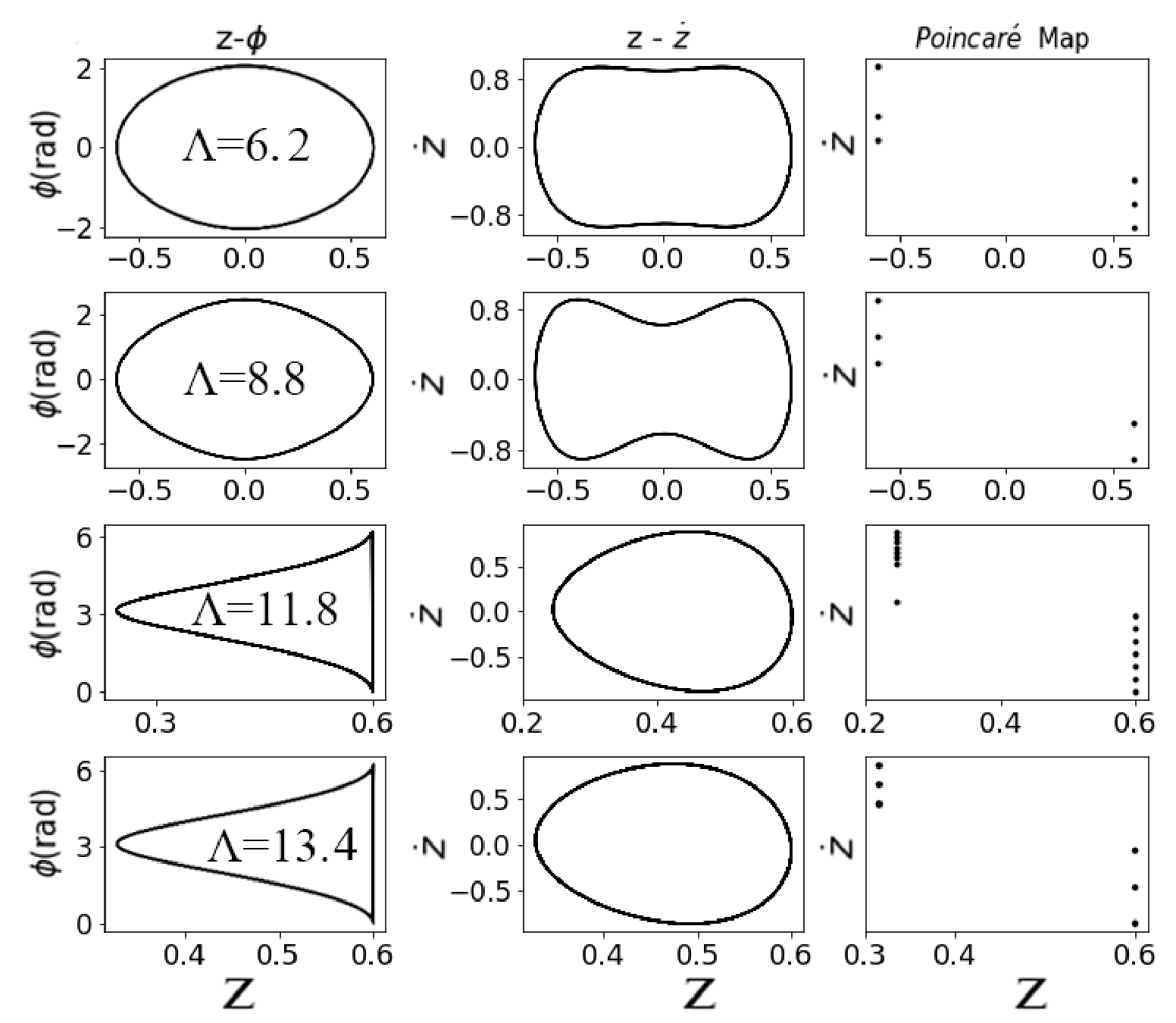}
\caption{{Poincar\'{e} section} for a symmetric double-well: Diagrams of z-$\phi$(left), z-{$\dot{z}$}(middle), Poincar\'{e} maps(right). From top to bottom $\Lambda=(6.2, 8.8, 11.8,13.4)$, respectively.}
\label{figureA1}
\end{figure}

\begin{figure}[h]
\centering
\includegraphics[width=8.2cm, height=8.2cm]{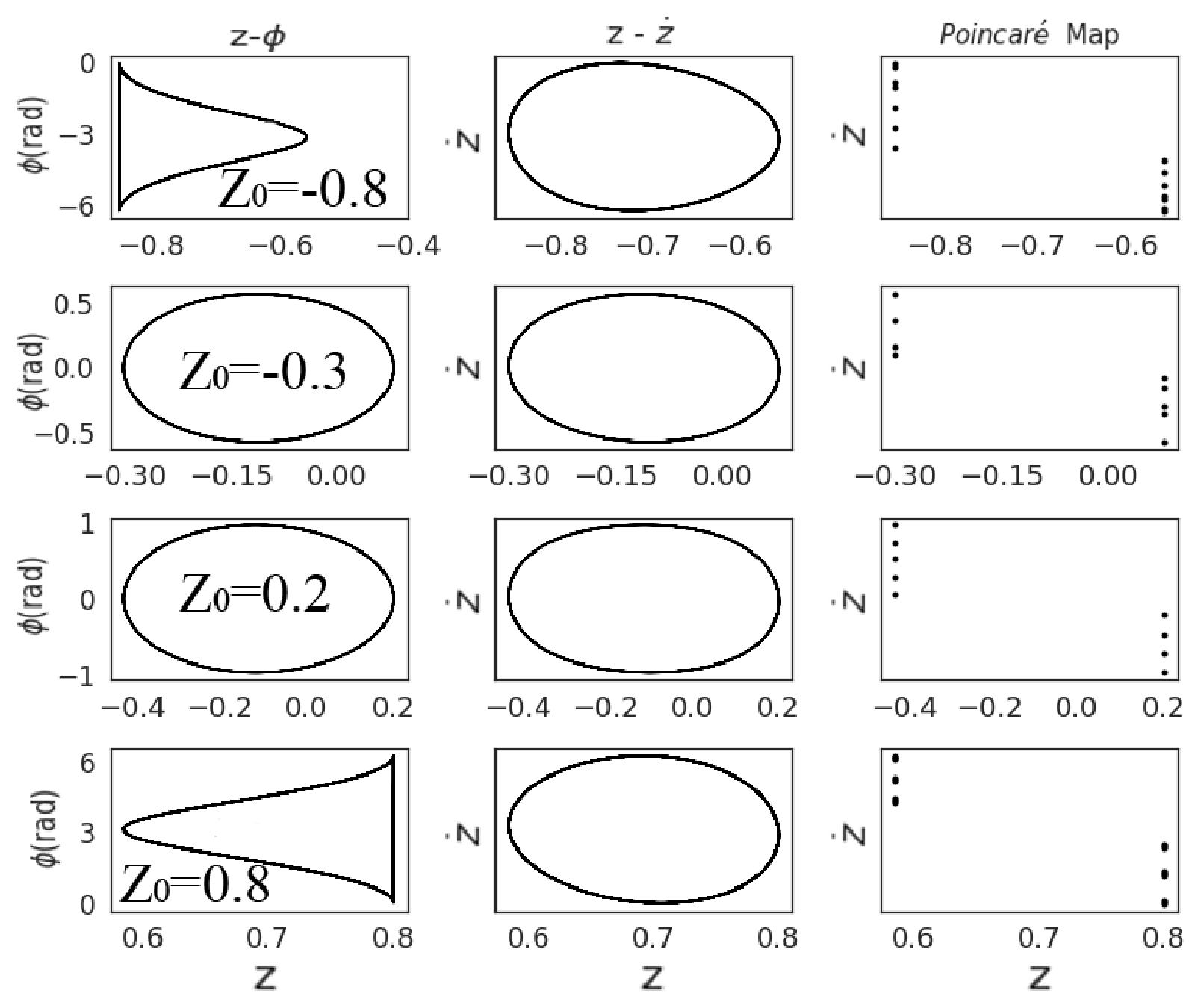}
\caption{{Poincar\'{e} section} for an asymmetric double-well: Diagrams of z-$\phi$(left), z-{$\dot{z}$}(middle), Poincar\'{e} maps(right). From top to bottom $z_0=(-0.8, -0.3, 0.2, 0.8)$, respectively.}
\label{figureA2}
\end{figure}

Some specific examples of Poincar\'{e} section for verifying the periodicity of population dynamics z(t), are given in
Figures \ref{figureA1} and \ref{figureA2}.

\nocite{*}

\end{document}